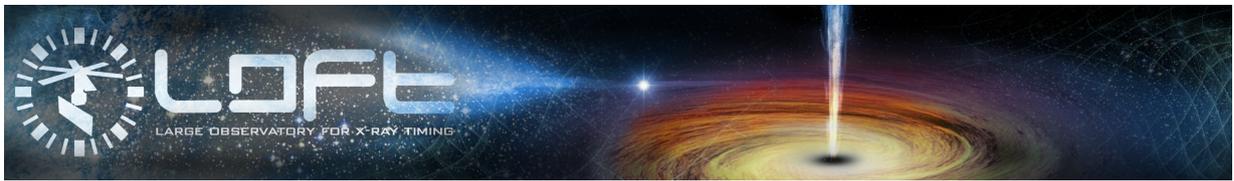

# Binary evolution with *LOFT*

## White Paper in Support of the Mission Concept of the Large Observatory for X-ray Timing

### Authors

T.J. Maccarone[1], R.A.M. Wijnands[2], N. Degenaar[3], A. Archibald[4], A. Watts[2], S. Vaughan[5], G. Wynn[5], G. Knevitt[5], W. Farr[6], N. Andersson[7], M. van der Klis[2], A. Patruno[8], T.M. Tauris[9]

[1] Texas Tech University, Lubbock, TX, USA
[2] University of Amsterdam, Amsterdam, Netherlands
[3] University of Cambridge, Cambridge, United Kingdom
[4] ASTRON, Dwingeloo, Netherlands
[5] University of Leicester, Leicester, United Kingdom
[6] University of Birmingham, Birmingham, United Kingdom
[7] University of Southampton, Southampton, United Kingdom
[8] Leiden University, the Netherlands
[9] AIfA (University of Bonn) and MPifR, Germany

**Preamble**

The Large Observatory for X-ray Timing, *LOFT*, is designed to perform fast X-ray timing and spectroscopy with uniquely large throughput (Feroci et al., 2014). *LOFT* focuses on two fundamental questions of ESA's Cosmic Vision Theme "Matter under extreme conditions": what is the equation of state of ultra-dense matter in neutron stars? Does matter orbiting close to the event horizon follow the predictions of general relativity? These goals are elaborated in the mission Yellow Book (http://sci.esa.int/loft/53447-loft-yellow-book/) describing the *LOFT* mission as proposed in M3, which closely resembles the *LOFT* mission now being proposed for M4.

The extensive assessment study of *LOFT* as ESA's M3 mission candidate demonstrates the high level of maturity and the technical feasibility of the mission, as well as the scientific importance of its unique core science goals. For this reason, the *LOFT* development has been continued, aiming at the new M4 launch opportunity, for which the M3 science goals have been confirmed. The unprecedentedly large effective area, large grasp, and spectroscopic capabilities of *LOFT*'s instruments make the mission capable of state-of-the-art science not only for its core science case, but also for many other open questions in astrophysics.

*LOFT*'s primary instrument is the Large Area Detector (LAD), a $8.5\,\mathrm{m}^2$ instrument operating in the 2–30 keV energy range, which will revolutionise studies of Galactic and extragalactic X-ray sources down to their fundamental time scales. The mission also features a Wide Field Monitor (WFM), which in the 2–50 keV range simultaneously observes more than a third of the sky at any time, detecting objects down to mCrab fluxes and providing data with excellent timing and spectral resolution. Additionally, the mission is equipped with an on-board alert system for the detection and rapid broadcasting to the ground of celestial bright and fast outbursts of X-rays (particularly, Gamma-ray Bursts).

This paper is one of twelve White Papers that illustrate the unique potential of *LOFT* as an X-ray observatory in a variety of astrophysical fields in addition to the core science.

# 1  Summary

The Large Observatory For X-ray Timing (*LOFT*) will present a range of new opportunities to study binary stellar evolution processes. Its Wide Field Monitor will be the most sensitive one deployed to date.

- LOFT will allow a census of Very Faint X-ray Binaries across a large fraction of the Galaxy. By establishing a substantial sample of such objects through the wide field coverage and good sensitivity of the Wide Field Monitor, it should be possible to make optical follow-up on enough of them to uncover their nature, which is still poorly understood.

- LOFT will allow us to determine whether the orbital period distribution of black hole X-ray binaries follows the predictions of binary evolution. At the present time, the severe selection biases against black hole X-ray binaries with orbital periods less than about 4 hours lead to a deficit of what it likely to be the largest group of systems. By enhancing the sample of objects for which masses can be estimated, LOFT should also help verify or refute the "mass gap" between neutron stars and black holes which has profound implications for understanding how supernovae actually explode.

- The process of spin-up of neutron stars in binaries can be probed in two manners: by using LAD measurements to expand the sample of neutron stars with good spin period measurements; and by using the combination of WFM and LAD to discover and study the population of objects that are in the act of transitioning from low mass X-ray binaries into millisecond radio pulsars.

# 2  Introduction

The evolution of binary stars is one of the most important and challenging problems in modern stellar astronomy. Binary evolution – even when restricted merely to the evolution of binaries with compact objects – touches on a range of key questions – the formation of Type Ia supernovae, of the formation of merging neutron stars and black holes, the rate of production of millisecond pulsars, and the rate of production of a variety of classes of stars with abundance anomalies. The challenges in understanding the populations of binary stars stem from the wide range of physics that goes into them. In addition to normal stellar evolutionary processes, the following must be understood: the common envelope stage of stellar evolution, the kicks applied to neutron stars and black holes at birth, the details of the process of supernova explosions, the angular momentum loss mechanisms that keep systems bound (i.e., gravitational radiation and magnetic braking), the initial parameter distributions of binary systems, and the distribution of the initial mass ratios of the stars in the binary systems.

Proper testing of theories of binary evolution must come from development of substantial samples of close binaries, and good estimates of their system parameters. Because most X-ray binaries, especially with black hole primaries, are transients, the ideal way to detect such objects is with all-sky monitoring. The *LOFT* Wide Field Monitor represents a major step forward in the capabilities of all-sky monitors relative to present monitors.

There are several major topics in compact binary populations toward which *LOFT* can make an essentialy contribution: the very faint X-ray transient problem; the orbital period distribution of black hole X-ray binaries; the "mass gap" between black holes and neutron stars that has been tentatively observed in existing data; and understanding the spin-up and spin-down processes in neutron stars by making better measurements of the neutron star spin distribution, observing the spin evolution of individual objects, and observing the transition objects which switch between accretion-powered and rotation-powered states.

## 3  Very faint X-ray binaries: general topics

Low-mass X-ray binaries accreting at a luminosity of $\gtrsim 10^{36}$ erg s$^{-1}$ have been extensively studied for decades, using many different observatories. Accordingly, their observational properties are very well characterized. However, with the advent of the current generation of sensitive X-ray instruments (e.g., *Chandra*, *XMM-Newton*, and *Swift*) a growing number of X-ray binaries have been found that exhibit peak luminosities that fall well below the typical sensitivity of X-ray monitoring instruments ($\simeq 10^{36}$ erg s$^{-1}$). Due to observational challenges, our knowledge of the properties of these *very-faint X-ray binaries* (VFXBs) remains limited.

VFXBs have both been seen as transient (exhibiting days–weeks long outbursts that peak below $\simeq 10^{36}$ erg s$^{-1}$) and as persistent X-ray sources (displaying a relatively steady luminosity of $\simeq 10^{34} \ldots 10^{36}$ erg s$^{-1}$ for several years). Currently about 3 dozen VFXBs are known in our Galaxy. Several of these have been established to harbor neutron stars through the detection of thermonuclear X-ray bursts (e.g., in 't Zand et al., 2005; Del Santo et al., 2007; Degenaar et al., 2010, 2013). However, there are also (short period) black holes expected among them (Armas Padilla et al., 2013; Corral-Santana et al., 2013; Maccarone & Patruno, 2013; Knevitt et al., 2014; Wijnands et al., 2014, see also Section 4).

Although the number of known VFXBs has been steadily growing in the past decade, their sub-luminous character is not yet fully understood. They may be ultra-compact X-ray binaries with orbital periods of <1.5 hr and small/evolved donor stars (e.g., King & Wijnands, 2006; in 't Zand et al., 2007; Heinke et al., 2015), or wind accretors (e.g., Maccarone & Patruno, 2013), although some appear to require a different explanation (e.g., Degenaar et al., 2010; Heinke et al., 2015). An alternative possibility is that some VFXBs harbor neutron stars with relatively strong magnetic fields (e.g., Wijnands, 2008; Patruno, 2010; Degenaar et al., 2014; Heinke et al., 2015) and are possibly related to the recently emerged class of "transitional objects", which switch between radio pulsar and X-ray binary manifestations (Papitto et al., 2013; Bogdanov et al., 2014; Archibald et al., 2014; Bassa et al., 2014). Strikingly, VFXBs significantly outnumber brighter X-ray binaries in the Galactic center region (e.g., Muno et al., 2005; Wijnands et al., 2006; Degenaar & Wijnands, 2009; Degenaar et al., 2012).

An important step in furthering our knowledge of VFXBs is to understand their binary parameters, which can be achieved using follow-up observations at optical and infrared wavelengths. However, such studies are hampered by the generally large distances (>8 kpc) and high absorption columns ($> 10^{22}$ cm$^{-2}$; $A_V > 40$ mag) of currently known VFXBs: most known systems are concentrated toward the Galactic center (this region has been intensively studied with sensitive X-ray instruments, allowing for the discovery of several VFXBs). *LOFT* would make an important contribution in taking the research of VFXBs to the next level. With the *LOFT*/WFM, VFXBs can be detected in large numbers in other parts of the Galaxy where the extinction is much less severe and optical/infrared follow-up observations are feasible. Moreover, mapping out the distribution of VFXBs will show whether their relative abundance compared to brighter X-ray binaries also holds in other parts of the Galaxy.

The *LOFT*/WFM will reach a daily sensitivity of $\simeq$3 mCrab in the Galactic Center region, and $\simeq$1 mCrab in other parts of the Galactic plane. Assuming an outburst duration of at least $\simeq$5 days (e.g., Degenaar & Wijnands, 2010), and coverage of the Galactic Plane and Galactic Center about 1/3 of the time, the detection sensitivity of VFXBs will be $\simeq$2.5 mCrab in the Galactic center, and $\simeq$0.7 mCrab in the plane. At a distance of 8 kpc, this would correspond to $\simeq 2 \times 10^{35}$ erg s$^{-1}$, which is a factor of 6-15 deeper than *RXTE*/ASM, MAXI, or *Swift*/BAT. In the past decade, about 1–2 new VFXBs have been found per year by all-sky monitors, but the other monitors were not sensitive enough to detect sources below $10^{36}$ erg s$^{-1}$ in the Galactic Center. The substantial increase in stellar mass within the horizon for VFXB detection that *LOFT* will provide means that it is likely to double the total population of VFXBs. The detection rate may, in fact, be even larger than this, based on expected orbital period distributions of black hole X-ray binaries (see Section 3).

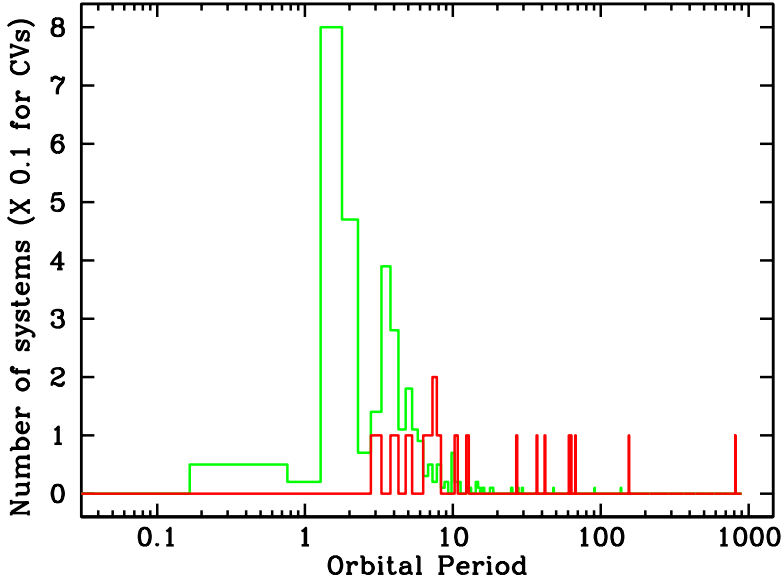

Figure 1: The orbital period distributions of dynamically confirmed black hole X-ray binaries (red curve), and cataclysmic variables, taken from the Ritter catalog (green curve, with the numbers multiplied by 0.1 to fit the two distributions well on the same scale). It is apparent that the distributions are very different from one another.

## 4 Orbital period distributions of black hole X-ray binaries

A basic understanding of how mass transferring binary stars evolve has existed for decades (e.g., Paczyński 1971), but a more detailed understanding has proved hard to develop. The evolution of low mass X-ray binaries should be quite similar to that of cataclysmic variables. Mass transfer from the lighter to the heavier object in a binary system pushes the binary toward larger separation and longer orbital period. In order for binary systems in which mass is transferred by Roche lobe overflow to remain in contact, either the donor star must be expanding, or the angular momentum must be transported out of the system. Expansion sufficient to maintain contact without additional angular momentum transport normally occurs for only the longest orbital period systems, which have subgiant or giant donor stars. Angular momentum transport from systems with relatively long orbital periods typically takes place via magnetic braking, while angular momentum transport from systems with shorter periods takes place via gravitational radiation.

The distributions of the orbital periods of the detected members of the two classes of systems are dramatically different. The cataclysmic variable show a strong peak to their orbital period distribution at about 2 hours, while the black holes show a flat period distribution, with *all* well-measured systems at periods longer than 2 hours.

Theoretical predictions for the orbital period distribution of black holes indicate that there should be a peak at a few hours, just as for the cataclysmic variables. King, Kolb & Burderi (1996) set out an analytic expression that provides a reasonable approximation to more detailed binary evolution calculations:

$$-\dot{M}_2 = \dot{M}_{MB} m_1^{-2/3} \hat{m}_2^{7/3} P_3^{5/3} + \dot{M}_{GR} m_1^{2/3} \hat{m}_2^2 P_3^{-2/3} \qquad (1)$$

where $M_2$ is the mass of the donor star, $\dot{M}_{MB}$ is the scale factor for the magnetic braking mass transfer rate, and is $2 \times 10^{-9}$ $M_\odot$ yr$^{-1}$, $m_1$ is the mass of the accretor, $\hat{m}_2$ is the ratio of the mass of the donor star to the mass of a Roche lobe filling main sequence star in the same period orbit, $P_3$ is the orbital period in units of 3 hours, and $\dot{m}_{GR}$ is the scale factor for the mass transfer rate due to gravitational radiation, and is $7.6 \times 10^{-11}$ $M_\odot$ yr$^{-1}$. The two terms become equal for periods of about 3 hours for black holes with masses of about 8–10 $M_\odot$.

Furthermore, the orbital period distribution should be *dominated* by the shorter period systems. To first approximation, the number of systems in existence at any given time should be proportional to the timescale of evolution at that period, which can be taken to be $M_2/\dot{m}_2$. Where magnetic braking dominates, the timescale of evolution is shortest for the longest period systems. Where gravitational radiation dominates, the timescale of evolution is *longest* for the longest period systems. Systems should thus pile up at orbital periods of about 3 hours.

These short period black hole X-ray binaries have been conspiuously absent in existing samples. The reason for this was recently explained well by Knevitt et al. (2014): these systems should all be transients, and their outbursts should be both relatively short in duration, and faint at their peaks (given the strong correlation between peak X-ray luminosity and orbital period – Shahbaz et al. 1998; Portegies Zwart et al. 2004; Wu et al. 2010). As a result, existing all-sky monitors will not detect the shorter period systems at large distances. In addition to the standard relationship between the peak mass transfer rate and the orbital period that comes from the density in the accretion disk needed to trigger the ionization instability, the short period black holes (below ~4 hour orbital periods) should be in radiatively inefficient states which should further reduce their peak luminosities (Wu et al. 2010; Knevitt et al. 2014). One additional short period system does exist which is likely to host a black hole, but which has not yet been dynamically confirmed – Swift J1357.2−0933 (Corral-Santana et al. 2013). Several additional longer period systems are also strong black hole candidates, but not dynamically confirmed.

The *LOFT* Wide Field Monitor should be capable of rectifying this situation and discovering appreciable numbers of these faint, fast X-ray transients. Let us consider a system with an orbital period of 3 hours. Cataclysmic variables clearly follow a relation where $L_{peak} \propto P_{orb}^{4/3}$ (Warner 1987; Maccarone 2013), and the peak values of $\dot{m}$ for X-ray binaries should be expected to do so as well. The low hard states are expected to be radiatively inefficient, with $L_X \propto \dot{m}^2$ below a state transition luminosity of about 2% of the Eddington limit (Maccarone 2003). As a result $L_{peak}$ for X-ray binaries is likely to scale as $P^{8/3}$ for periods short enough that a soft state is never reached.

If we take the system XTE J1118+480 as a template, we can make an estimate of the distance out to which the *LOFT* WFM can be expected to detect short period black hole X-ray binary transients. This system has an orbital period of 4.1 hours and is located at a distance of 1.8 kpc. Its peak 3–200 keV X-ray luminosity was about $1.5 \times 10^{36}$ erg s$^{-1}$ (Wu et al. 2010), and its original outburst peak lasted about 5 days, followed quickly by a second outburst which lasted about 100 days near its peak luminosity (e.g., Wood et al. 2001). Systems at 3 hour period should then peak at about $6.5 \times 10^{35}$ erg s$^{-1}$, while systems at 2 hours orbital periods should peak at about $2.2 \times 10^{35}$ erg s$^{-1}$. The peak flux was only 40 mCrab, despite the source being very near to us. The *RXTE* ASM's sensitivity limit was about 10 mCrab, so that even at a distance of 4 kpc, this source would likely have gone undetected; similarly, even at such a short distance, systems with orbital periods of 2 hours would be undetectable.

In recent years, three new transients have been seen which are strong black hole candidates, and which have orbital periods less than 4 hours – Swift J1753.5−0127, MAXI J1659−152 and Swift J1357.2−0933. It is notable that they are located 12, 17 and 50 degrees out of the Galactic Plane, respectively. While MAXI and *Swift* provide better sensitivity in general than *RXTE*'s All Sky Monitor, the combination of having very good angular resolution for a coded mask, and a set of modules with modest fields of view give the *LOFT* WFM dramatically better sensitivity in the most crowded parts of the sky.

Systems with orbital periods of 3 hours should be detectable out to a distance of 3.3 kpc in the Galactic Center region, and about 6 kpc in other parts of the Galactic Plane, while systems with periods of 2 hours should be detectable out to distances of about 2 and about 5 kpc, respectively.

The survey region should then be approximately one third of the stellar mass in the Galaxy for systems with orbital periods of 3 hours. The present survey region for such objects, out to a distance of about 2.5 kpc, includes only about 3% of the Milky Way's stellar mass.

Next, we can make estimates of the number of these systems we should detect. The timescale for a black hole to evolve from an orbital period of 10 hours to a period of 3 hours should be about 200 Myrs. The timescale to evolve from 3 hours to 2 hours should be about 2.4 Gyr. We therefore expect 12 objects at 2–3 hour periods[1] for every object in the 3–10 hour period range. Knevitt et al. (2014) find that the expected rate of outbursts should be nearly constant for orbital periods between about 1 and 20 hours. *RXTE* detected ~1 new black hole transient per year in the 3–10 hour period range. *LOFT* WFM should thus detect about 3–4 new transients per year in the short orbital period range of the period distribution; the number may be even higher if a substantial fraction of initial donor star masses in BH LMXBs are below 1 $M_\odot$, or if a substantial fraction of the outburst durations are 50 days rather than 5 days.

---

[1]This period range represents the "period gap" for cataclysmic variables, where few systems are found as stars initially become fully convective. It should be dramatically less important for black holes than for cataclysmic variables because of the heavier accretors –

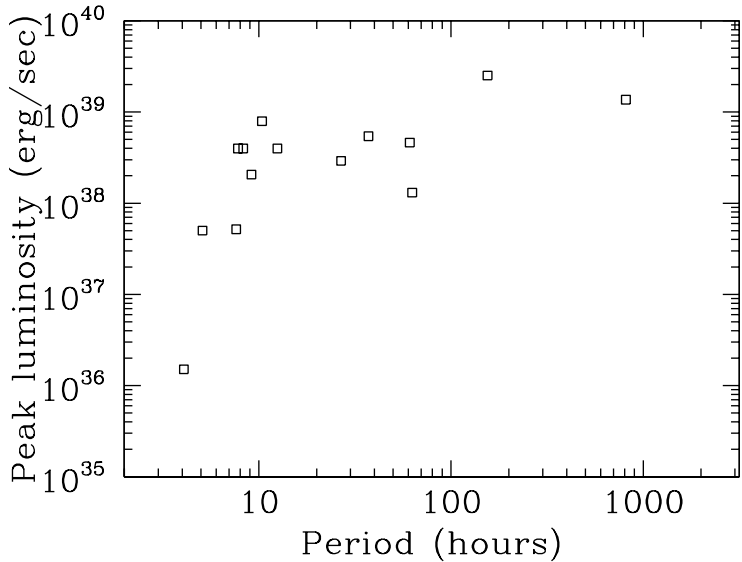

Figure 2: Peak luminosity plotted versus orbital period for the black hole X-ray transients seen with RXTE, with data taken from Wu et al. 2010. Pre-RXTE systems have been added, with data taken from Garcia et al. (2003). Similar relations are also seen for cataclysmic variables (Warner 1987; Maccarone 2015) and accreting neutron stars (Wu et al. 2010).

## 5 Expanding the sample of black hole masses to probe the mass gap

At the present time, only about 20 black holes have precisely estimated masses. This sample already shows that the mass distribution of black holes appears to be sharply peaked at about 8 $M_\odot$, with an apparent substantial gap between the maximum mass for a neutron star (even if one allows that to go to 3 $M_\odot$) and the minimum mass for a black hole. Özel et al. (2010) find that the black hole mass distribution can be well described as a Gaussian with mean 7.8 $M_\odot$ and a dispersion of 1.2 $M_\odot$, while Farr et al. (2011) consider a broader range of mass distributions but find that for more reasonable sets of assumptions, fewer than 1% of black holes should be less than 4 $M_\odot$. Additional, but somewhat indirect, indications of a similar "mass gap" come from studies of the nearby galaxy NGC 5128, which shows two tracks of thermally emitting sources in a temperature-luminosity diagram with a gap between them (Burke et al. 2013).

Some attempts have been made to account for some possible selection effects in the existing sample of black hole X-ray binaries with estimated masses. Özel et al (2010) showed that the mass gap cannot be the result of selection effects that depend solely on the mass of the black hole. However, neither they, nor Farr et al. (2011) considered the possibility of a relation between the orbital period and the black hole mass, something which is at least tentatively suggested from observational data, especially if one includes high mass X-ray binaries (e.g., Lee et al. 2002). If the mass-period relation hypothesis is correct, building up a sample of X-ray binaries with short orbital periods may fill in the mass gap.

At the present time, the explanation for the mass gap which has gained the most traction is that the supernovae explosions producing black holes must be "fast" – i.e., driven by instabilities with growth timescales of ~ 10 milliseconds happening very shortly after the start of core collapse, so that the explosion takes place within

---

King et al. 1997.

a few hundred milliseconds of the core collapse. Such a constraint provides vital feedback into modelling of the still poorly understood process of core collapse supernovae (Belczynski et al. 2012).

What is needed to move forward with determing whether, and how, the black hole mass distribution constrains supernova mechanisms is a larger sample of well-estimated black hole masses. The *LOFT* WFM should roughly double the sample of known black hole candidates in the Galaxy. By 2025, new small telescopes covering the whole sky with high cadence (Law et al. 2014) should make discovery of X-ray binaries' optical counterparts in outburst straightforward, and should provide measurements of superhump periods which can be used to estimate mass ratios in mass transferring binaries (e.g., Patterson et al. 2005). The 30-m class of telescopes working in the optical and infrared should have the capabilities to measure mass functions even for systems with faint M-dwarf donors at 5 kpc distances. *LOFT* is what is needed to develop large samples of black hole masses in the context of the multi-wavelength astronomy landscape that will be available in the mid-2020's.

## 6  Neutron star spin distributions

*LOFT*, with its exquisite sensitivity to pulsations, offers a unique opportunity to fully characterise the spin distribution of neutron stars. The spin distribution provides a guide to the torque mechanisms in operation and the moment of inertia. However it is at present poorly understood. The evolution of a neutron star born in a binary system (assuming that it survives the initial supernova) can follow various routes (Podsiadlowski et al. 2002; Tauris & van den Heuvel 2006; Lorimer, 2008). This may include a period of accretion onto one of the neutron stars, from a main sequence or white dwarf companion, during which time the system may be visible in X-rays. Spin-up due to such accretion is the basis for the recycling scenario that is thought to explain the formation of the Millisecond Radio Pulsars (MSPs – Alpar et al. 1982; Radhakrishnan & Srinivasan, 1982; Bhattacharya & van den Heuvel, 1991). Several studies have attempted to link the properties of the MSPs to those of the neutron stars in Low Mass X-ray Binaries (LMXBs), the accreting systems that are supposed to be the progenitors of MSPs in the recycling scenario. The discovery of the first accreting millisecond X-ray pulsar by Wijnands & van der Klis (1998), and the recent detection of a transitional objects that switch from radio to X-ray sources (Archibald, et al., 2009; Papitto, et al., 2013) seems to confirm this picture. However detailed comparisons of the properties (such as spin and orbital periods) of neutron star LMXBs and the radio MSPs reveal discrepancies between the predictions of current evolutionary models and observations (Hessels, 2008; Kiziltan & Thorsett 2009; Tauris et al. 2012). This suggests shortcomings in our understanding of mass transfer, magnetic field decay, and accretion torques. It may even be necessary to consider alternative formation routes for the MSPs (Knigge et al. 2011). It is not clear that the spin distribution of the MSPs provides a good guide to the spin distribution of the accreting sources: and based on lifetimes and accretion torque estimates it is quite feasible that accreting NS may achieve higher spins than those measured for the MSPs (Cook, Shapiro, & Teukolsky, 1994).

*LOFT* 's capabilities are well suited to discover many more neutron star spins. It is now known that accretion-powered pulsations from these sources can show up intermittently (Galloway et al. 2007; Casella et al. 2008; Altamirano et al. 2008). Theory predicts intermittent episodes of channeled accretion onto weakly magnetized neutron stars in high accretion rate systems (Romanova et al. 2008), and intermittency is also likely in systems where the system is close to alignment (Lamb, et al., 2009). So, sensitivity to brief pulsation trains is key. Using the LAD, *LOFT* will detect 100 s duration pulse trains down to an amplitude of 0.4% rms for a 100 mCrab source ($5\sigma$). This time interval matches the short duration of intermittent pulsations such as observed with *RXTE* (Casella et al. 2008). *RXTE* needed 15 times as long to reach the same sensitivity, so that 100 s pulse trains were severely diluted; longer pulse trains suffered from Doppler smearing. Weak pulsations are also expected in systems where magnetic field evolution caused by accretion has driven the system towards alignment (Ruderman, 1991). Searches with *LOFT* for weak (rather than intermittent) accretion-powered pulsations will use sophisticated search techniques such as those used to find the *Fermi* pulsars (Atwood et al. 2006; Abdo, et al., 2009; Messenger, 2011; Pletsch et al., 2012), to compensate for orbital Doppler smearing. Even at current computational capabilities, the *LOFT* core programme observations would yield $5\sigma$ multi-trial pulsation sensitivities of ∼0.03-0.003% (rms) in bright NS (> 100 mCrab, better for the very brightest sources) and ∼0.2-0.04% in faint NS (10-100 mCrab), taking into account expected observing times and prior knowledge of the orbits. *LOFT* will also be able to measure spins using thermonuclear burst oscillations (see Watts 2012 for a review). *LOFT* will be able to detect oscillations in individual Type I X-ray bursts to amplitudes of 1.1% (2.7%) rms in the burst tail (rise); by stacking bursts sensitivity improves. A single measurement of an extremely rapid spin provides a simple and very clean constraint on the equation of state (EOS), part of the core *LOFT* science case. It is possible, however, that with more spin detections we instead establish the existence of a pile-up in the distribution at a rotation rate well below the mass shedding limit (Bildsten, 1998). Discovery of an effect that prevents accretion-induced spin-up to reach this limit would be an important result and would suggest gravitational wave or magnetic braking mechanisms dominate neutron star spin histories.

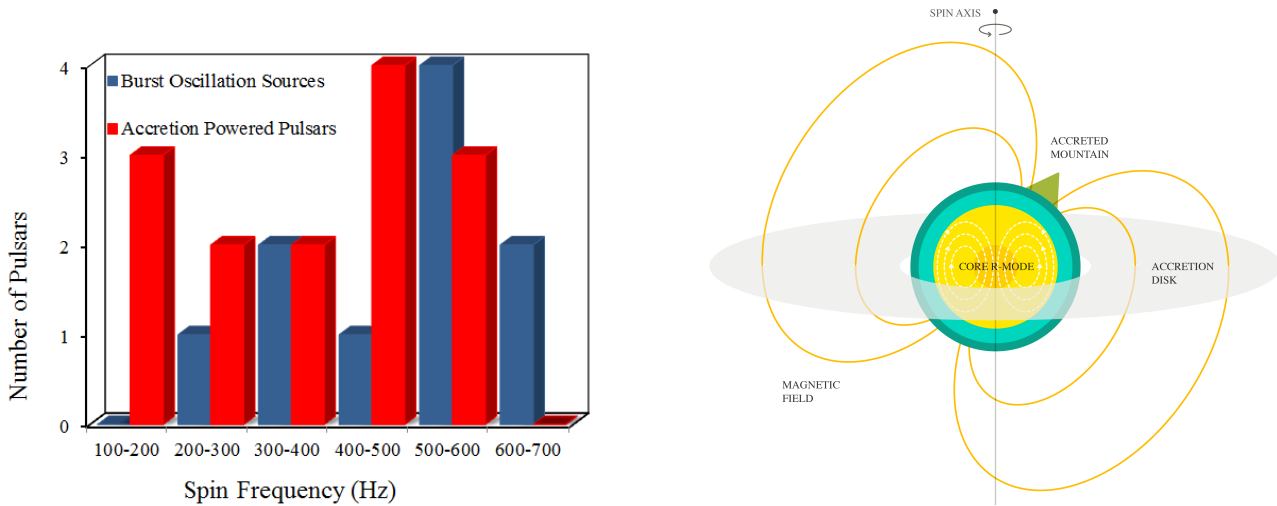

Figure 3: *Left:* The spin distribution for the fastest known accreting neutron stars (not including the three transitional pulsars). *Right:* The various torque mechanisms operating on an accreting neutron star. In addition to the accretion torque, which might involve magnetic channelling, there are gravitational wave torques such as caused by crustal mountains or core r-modes that depend sensitively on the dense matter EOS.

The spin evolution of accreting neutron stars (Fig. 3) is intimately connected to details of the internal physics, such as core r-mode oscillations (Ho et al. 2011) or crustal deformations such as "mountains" (Johnson-McDaniel & Owen, 2013) that generate gravitational waves, the response of the superfluid, and the moment of inertia. Disentangling the effects of magnetic accretion torques will be important in this effort (Ghosh & Lamb, 1978; Andersson et al. 2005). Excitingly, this will also enable us probe the physics of the weak interaction at high densities, since weak interactions control the viscous processes that are an integral part of the gravitational wave torque mechanisms (Alford, Mahmoodifar, & Schwenzer, 2012).

## 7 Millisecond pulsar/X-ray binary transition objects

Light has recently been shed on the evolution from LMXBs into MSPs by the discovery of three systems that transition between the two states: PSR J1023+0038, XSS J12270−4859, and IGR J18245−245 (M28I). The transitional nature of these systems promises to shed light on the end of accretion and the birth of an MSP, but the ability to observe these systems in two states means that information only available in one state can shed light on the physics occurring in the other (for example, in PSR J1023+0038, observations in the MSP state have revealed not just the spin period and magnetic field but the distance, geometry, and pulsar and companion masses; Deller et al. 2012, Archibald et al. 2013). Currently, though, the scientific possibilities are limited by the small number of known transitional systems and the rarity of their transitions. *LOFT* offers the possibility of finding more such systems blindly as well as detecting the transitions of systems (in particular the "redback" and "black widow" MSPs) suspected to be transitional. Further, *LOFT*'s sensitivity to pulsations and short-term variability should allow detailed study of new and existing transitional systems in their LMXB states given the radio pulse ephemerides already available.

As LMXBs, two of the known transitional systems (PSR J1023+0038 and XSS J12270−4859) have sustained luminosities of roughly $3 \times 10^{33}$ erg s$^{-1}$, although their variability seems to take the form of switching between three modes: relatively stable low ($\sim 5 \times 10^{32}$ erg s$^{-1}$) and high ($3 \times 10^{33}$ erg s$^{-1}$) modes, and a brighter flaring mode ($1–3 \times 10^{34}$ erg s$^{-1}$). These luminosities are typical of quiescent LMXBs, but both systems have shown evidence for accretion-powered pulsations in the high mode (but neither of the others; Archibald et al. 2014, Papitto et al. 2014). These systems are therefore examples of very faint LMXBs as discussed above, but the observation of a MSP state makes these systems particularly interesting. The third known transitional system, M28I, went into full-fledged outburst, but also spent time in such a variable mode-switching state.

The presence of the different luminosity modes, and the rapid switching between them, is an outstanding puzzle. Archibald et al. (2014) suggested that the mode switching might have to do with transitions between "trapped disk" and "propeller" modes in the inner accretion disk region, but no existing theoretical model can explain all aspects of the observed data. Deller et al. (2014) observed variable flat-spectrum continuum emission from PSR J1023+0038 like that seen from jets, but no data was available to test for correlated variability between the radio and X-ray. *LOFT* has the sensitivity to make a flux measurement on PSR J1023+0038 good to 10% within 16 seconds, so a detailed study of the light curves should be possible. *LOFT*'s sensitivity to pulsations (it would detect the pulsations in PSR J1023+0038's high state in about 300 s), particularly in combination with radio-derived pulse timing ephemerides, will also allow testing whether channelled accretion onto the surface occurs in the low and flare modes.

The ability to detect pulsations from transitional systems while they are accreting offers the possibility of applying the techniques of pulsar timing to these systems. Models of these systems that involve an active radio pulsar require spin-down of a roughly known magnitude, while accretion onto the surface leads to spin-up torques and ejection from the system requires spin-down torques. Further, in their MSP state these systems exhibit apparently random orbital period variations (presumably linked to motions within the companion; Archibald et al. 2013). If these variations are due to the mechanism of Applegate and Shaham (1994), where they are due to changes in the companion shape, they may well be linked to the Roche lobe overflow, and measurements of them may shed light on the reason these systems switch between LMXB and MSP states. Unlike most accreting millisecond X-ray pulsars, the transitional systems seem to have stable X-ray pulse profiles (Archibald et al. 2014), so we expect phase shifts to be due only to pulsar timing effects. Currently only *XMM/Newton* is capable of detecting pulsations from the fastest systems, and limitations with its onboard clock (Martin-Carrillo et al. 2012) and scheduling complexities pose serious difficulties in carrying out timing measurements. In contrast, *LOFT* should be able to obtain a pulse phase measurement in only a few hundred seconds, so a series of short observations might permit phase-coherent timing.

With only three known systems, one of which is at a distance of 5.5 kpc, generalizations about transitional systems are difficult. Further, only one of the systems is currently in an LMXB state, so even comparisons between systems are challenging. A key priority in their study, therefore, is to discover more transitional systems and to detect when they transition so as to trigger multiwavelength follow-up. In their MSP state, the known transitional systems appear to be "redbacks", binary MSPs with main-sequence like companions (abnormal in surface abundance and structure, but which are, like main sequence stars, burning hydrogen) and evidence for interaction in the form of radio eclipses and orbital period variations. Eighteen such MSPs are now known (and more are being found rapidly, thanks in part to the combination of the Fermi gamma-ray telescope and ground-based radio follow-up); they are promising candidates for undergoing a transition to an LMXB state. There are also 35 "black widow" systems with lower-mass companions that show similar signs of interaction. The WFM of *LOFT* would easily detect X-ray emission from PSR J1023+0038 in the flare mode. Other transitioning sources may be at greater or lesser distances than the 1.37 kpc for PSR J1023+0038, but the flare mode is variable in brightness, and M28I reached a peak luminosity over $10^{37}$ erg s$^{-1}$, so we should expect most nearby transitioning systems to be detectable by the WFM. Follow-up (or periodic monitoring) with the LAD would immediately reveal whether a candidate system was in the LMXB state or the MSP state. Of course, the search for VFXBs may turn up some that are known as MSPs, or radio follow-up of VFXBs may reveal that some become active as MSPs. A large sample of transitioning systems and detailed observation of their transitions would certainly shed light on this newly-identified stage of binary evolution.

## 8 Other topics in binary evolution and populations

We briefly note that *LOFT* should make an impact on several other topcs in the evolution of compact binaries. Notably, the accreting millisecond X-ray pulsars are predominantly short orbital period binaries with low peak accretion rates. *LOFT* should help to expand the sample of them, with a goal of leading into a better understanding of both the spin-up and spin-down of neutron stars in binaries and of how X-ray binaries evolve into millisecond radio pulsars. Topics covered in other *LOFT* White Papers also have substantial impact on our understanding of binary stellar evolution – e.g., understanding the total mass loss rate in disk winds is fundamentally important for understanding whether mass transfer in low mass X-ray binaries is largely conservative or non-conservative; if high mean mass loss rates can be verified to be the rule, this may help explain why so few millisecond pulsars are found at masses substantially larger than 1.4 $M_\odot$.